\documentclass[final,5p,times,twocolumn]{elsarticle}

\usepackage{amsmath}
\usepackage{amssymb}
\usepackage{graphicx}
\usepackage{color}
\usepackage{xspace}

\usepackage[mathscr,scaled=1.15]{urwchancal}
\DeclareFontFamily{OT1}{pzc}{}
\DeclareFontShape{OT1}{pzc}{m}{it}%
{<-> s * [1.15] pzcmi7t}{}
\DeclareMathAlphabet{\mathpzc}{OT1}{pzc}{m}{it}

\definecolor{purple}{rgb}{0.5,0,0.5}
\definecolor{blue}{rgb}{0.0,0,0.9}
\definecolor{prdblue}{rgb}{0.133,0.118,0.498}

\biboptions{sort&compress}

\journal{Physics Letters B}

\newcommand{\be}{\begin{equation}}
\newcommand{\bea}{\begin{eqnarray}}
\newcommand{\ee}{\end{equation}}
\newcommand{\eea}{\end{eqnarray}}

\def\spr{\!\cdot\!}
 \def\s#1{{\scriptscriptstyle #1}}

\def\1eq#1{Eq.~(\ref{#1})}

\def\2eqs#1#2{Eqs.~(\ref{#1}) and~(\ref{#2})}
\def\3eqs#1#2#3{Eqs.~(\ref{#1}),~(\ref{#2}) and~(\ref{#3})}

\def\Nft{(2,0)}
\def\Nftpopo{(2,1,1)}
\def\nf{n_\s{\mathrm{f}}}

\def\aT{\alpha_\s{\mathrm{T}}}

\begin{document}

\begin{frontmatter}

\title{Scale-setting, flavour dependence and chiral symmetry restoration}

\author[ECT]{Daniele Binosi}
\author[ANL]{Craig D.~Roberts}
\author[UH]{Jose Rodr\'{\i}guez-Quintero}

\address[ECT]{European Centre for Theoretical Studies in Nuclear Physics
and Related Areas (ECT$^\ast$) and Fondazione Bruno Kessler\\ Villa Tambosi, Strada delle Tabarelle 286, I-38123 Villazzano (TN), Italy}
\address[ANL]{Physics Division, Argonne National Laboratory, Argonne, Illinois 60439, USA}

\address[UH]{Department of Integrated Sciences;
University of Huelva, E-21071 Huelva, Spain.}

\date{28 October 2016}

\begin{abstract}
We determine the flavour dependence of the renormalisation-group-invariant running interaction through judicious use of both unquenched Dyson-Schwinger equation and lattice results for QCD's gauge-sector two-point functions.  An important step is the introduction of a physical scale setting procedure that enables a realistic expression of the effect of different numbers of active quark flavours on the interaction.  Using this running interaction in concert with a well constrained class of dressed--gluon-quark vertices, we estimate the critical number of active lighter-quarks above which dynamical chiral symmetry breaking becomes impossible:
$n_f^{\rm cr}\approx 9$;
and hence in whose neighbourhood QCD is plausibly a conformal theory.
\end{abstract}

\end{frontmatter}


\noindent\textbf{1.$\;$Introduction}.
The last decade has seen the gauge sector of QCD provide important clues to some of the many puzzles encountered in the quest to understand the infrared (IR) dynamics of strongly-coupled theories.  Of particular interest is the feature that the gluon propagator saturates at infrared momenta, \emph{i.e}.\ $\Delta(k^2\simeq 0) = 1/m_g^2$, $m_g\simeq 0.5$\,GeV \cite{Cucchieri:2007md, Cucchieri:2007rg, Aguilar:2008xm, Dudal:2008sp, Bogolubsky:2009dc, Aguilar:2012rz}, which entails that the long-range propagation characteristics of gluons are dramatically affected by their self-interactions.  A similar feature is expressed in the dressed-quark propagator \cite{Bhagwat:2003vw, Bowman:2005vx, Bhagwat:2006tu}; and, hence, it is now known that the Schwinger functions of both these elementary coloured excitations violate reflection positivity, a sufficient condition for confinement \cite{Munczek:1983dx, Stingl:1985hx, Krein:1990sf, Burden:1991gd, Hawes:1993ef, Maris:1994ux, Bhagwat:2002tx, Roberts:2007ji, Bashir:2013zha, Qin:2013ufa, Lowdon:2015fig, Lucha:2016vte}.  A consistent picture is thus beginning to emerge: strong dynamics generates IR cutoffs in QCD so that long-wavelength ($\lambda \gtrsim 2/m_g\sim 1\,$fm) coloured-modes decouple and their role in hadron physics is superseded by interactions between light-hadrons \cite{Brodsky:2008be, Brodsky:2012ku, Binosi:2014aea, Horn:2016rip}.

The so-called ghosts, which represent the other component of the gauge sector, have also been thoroughly studied.  In this case it is their dressing function ({\it viz}.\ propagator$\,\times\,$momentum-squared) that saturates in the IR.  Consequently, even non-perturbatively, ghosts remain massless, being described by a simple $1/q^2$ propagator (up to logarithms) \cite{Aguilar:2008xm, Boucaud:2008ky, Cucchieri:2008fc, Dudal:2008sp, Bogolubsky:2009dc}.

It has steadily become clearer that a veracious expression of these features of gauge-sector dynamics is critical to the success of any continuum study of QCD and hadron observables.  This has, \emph{e.g}.\ recently enabled unification \cite{Binosi:2014aea}  of the top-down approach to determining the quark-antiquark scattering kernel directly from analyses of gauge-sector dynamics \cite{Aguilar:2009nf, Aguilar:2010gm} with the bottom-up approach, which uses a sophisticated, non-perturbative, symmetry-preserving truncation of matter-sector bound-state equations in order to construct a solution to the same problem via a comparison with empirical data \cite{Chang:2009zb, Chang:2010hb, Chang:2011ei, Chang:2013pq}.

In order to maintain momentum following that stride toward a continuum framework capable of providing \emph{bona fide} predictions of observables in continuum-QCD, herein we address additional, crucial issues.  Namely, how does the renormalisation-group-invariant (RGI) running interaction depend on the number of active quark flavours, $n_f$, and how best may one use results from lattice-regularised QCD (lQCD) to provide an answer?  The solutions to these puzzles will expand understanding of, \emph{inter alia}, confinement and dynamical chiral symmetry breaking (DCSB), and the $n_f$-dependence of observable hadron properties; and inform attempts to develop models for new physics based upon non-Abelian gauge theories (see, e.g.\ Refs.\,\cite{Appelquist:1996dq, Sannino:2009za, Appelquist:2009ka, Hayakawa:2010yn, Cheng:2013eu, Aoki:2013xza, DeGrand:2015zxa}).

\smallskip

\begin{figure*}[!t]
\begin{center}
\hspace{-0.75cm}
	\includegraphics[width=0.475\linewidth]{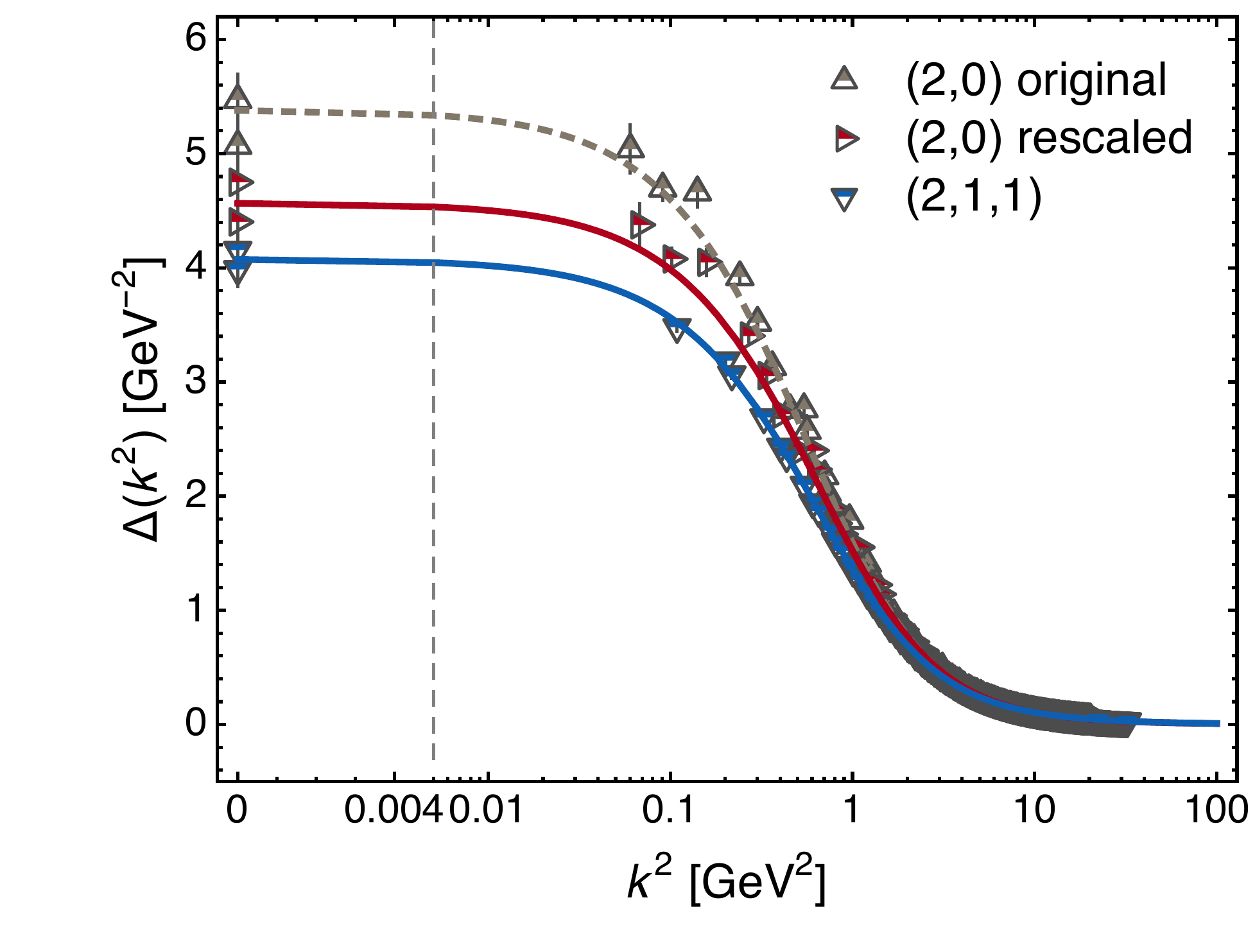}\hspace{0.5cm}
	\includegraphics[width=0.475\linewidth]{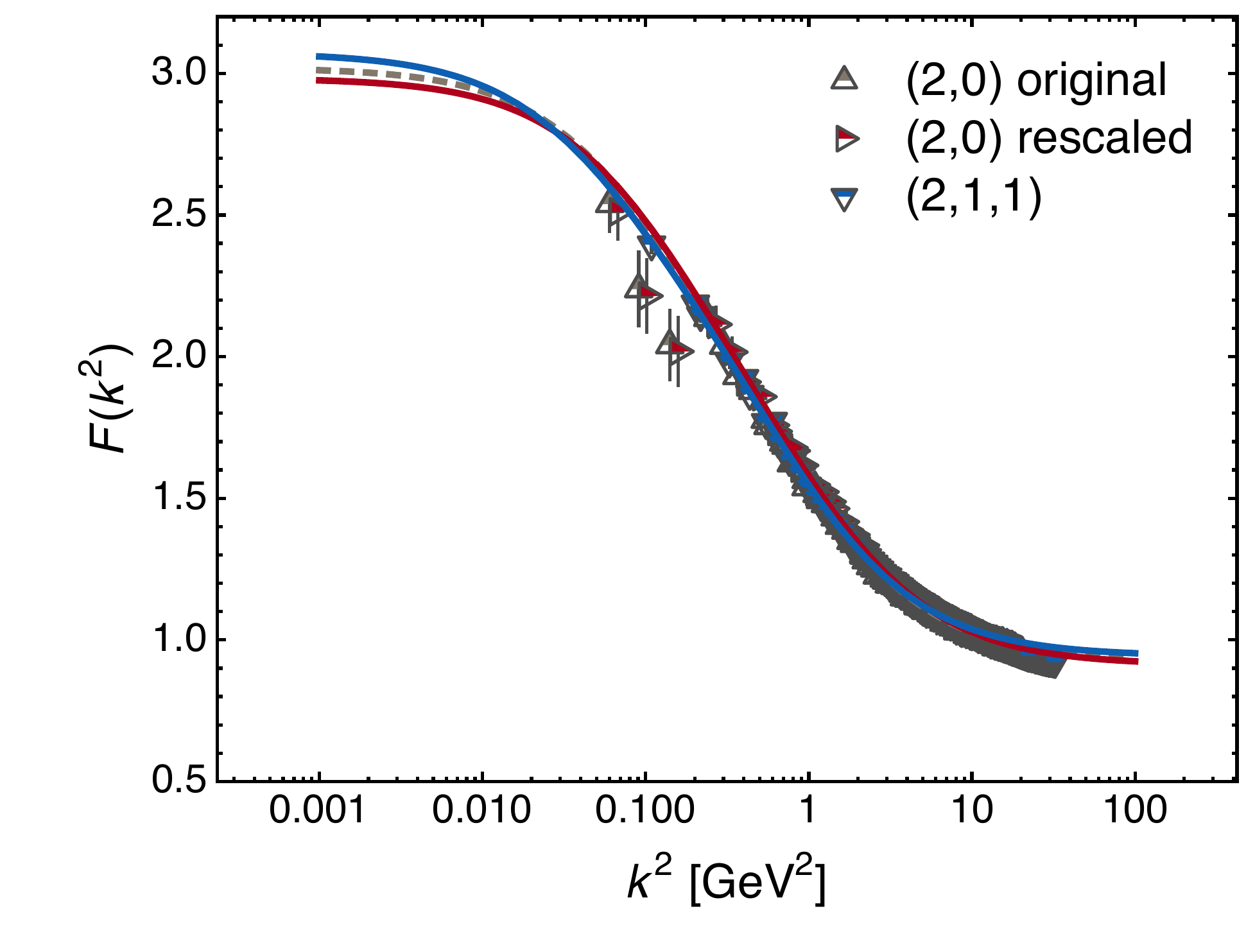}
\end{center}
\vspace{-.75cm}	
	\caption{\label{Delta-F} Functions characterising the gluon (left panel) and ghost propagators (right panel) obtained from numerical simulations of lQCD with $n_f=\Nft$ and $n_f=\Nftpopo$ \cite{Ayala:2012pb}.
Regarding $\Delta(k^2)$, the curves represent a fit, whereas for the ghost dressing function they depict the solution of the corresponding DSE.
For $n_f=\Nft$ we plot both the original lQCD results and the values obtained after rescaling as described in association with Eqs.\,\eqref{res-cond}\,--\,\eqref{rscale}.   Notably, the ghost is hardly affected by rescaling.
In the left panel the x-axis scale is linear to the left of the vertical dashed line and logarithmic otherwise, an artifice which  enables us to show the appearance of a gluon mass-scale at IR momenta.}
\end{figure*}

\noindent\textbf{2.$\;$Gap equation's kernel}.
A basic link between gauge-sector dynamics and QCD observables is the gap equation:
\begin{subequations}
\label{gendseN}
\begin{align}
S^{-1}(p) & = i\gamma\cdot p \, A(p^2) + B(p^2) \\
& = Z_2 \,(i\gamma\cdot p + m^{\rm bm}) + \Sigma(p)\,,\\
\Sigma(p)& =  Z_1 \int^\Lambda_{dq}\!\! g^2 D_{\mu\nu}(k)\frac{\lambda^a}{2}\gamma_\mu S(q) \frac{\lambda^a}{2}\Gamma_\nu(q,p) ,
\end{align}
\end{subequations}
where $D_{\mu\nu}(k=p-q)=\Delta(k^2) T_{\mu\nu}(k)$, $T_{\mu\nu}(k)=\delta_{\mu\nu}-k_\mu k_\nu/k^2$, is the gluon propagator in Landau gauge;\footnote{
Landau gauge is typically used because it is, \emph{inter alia} \protect\cite{Bashir:2008fk, Bashir:2009fv, Raya:2013ina}: a fixed point of the renormalisation group; that gauge for which sensitivity to model-dependent differences between \emph{Ans\"atze} for the gluon-quark vertex are least noticeable; and a covariant gauge, which is readily implemented in simulations of lattice-regularised QCD. Importantly, gauge covariance of Schwinger functions obviates any question about the gauge dependence of gauge invariant quantities.}
$\Gamma_\nu$, the quark-gluon vertex; $\int^\Lambda_{dq}$, a symbol representing a Poincar\'e invariant regularisation of the four-dimensional integral, with $\Lambda$ the regularisation mass-scale; $m^{\rm bm}(\Lambda)$, the current-quark bare mass; and $Z_{1,2}(\zeta^2,\Lambda^2)$, respectively, the vertex and quark wave-function renormalisation constants, with $\zeta$ the renormalisation point, which is usually $\zeta=\zeta_4:=3.61\,$GeV herein.

Whether or not DCSB and, arguably, confinement, too, emerge in the Standard Model is decided by the structure of the gap equation's kernel; and the interaction which unifies the top-down and bottom-up approaches to QCD's gauge sector may be expressed \cite{Binosi:2014aea}:
\begin{subequations}
\label{dkernel}
\begin{align}
	Z_1 g^2 D_{\mu\nu}(k) & = 4 \pi Z_2 \widehat{d}(k^2)  T_{\mu\nu}(k), \\
	\mathpzc{I}(k^2) := k^2 \widehat{d}(k^2) &= \frac{\aT(k^2)}{[1-L(k^2)F(k^2)]^2}.
	\label{iKer}
\end{align}
\end{subequations}
Here $\widehat{d}(k^2)$ is the renormalisation-group-invariant (RGI) function discussed in Ref.\,\cite{Aguilar:2009nf}, which arises naturally when combining the pinch technique \cite{Cornwall:1981zr, Cornwall:1989gv, Pilaftsis:1996fh, Binosi:2002ft, Binosi:2003rr, Binosi:2009qm} and background field method \cite{Abbott:1980hw, Abbott:1981ke} in analysing gauge-sector dynamics; $\aT$ is the ``Taylor coupling'' \cite{Blossier:2011tf, Blossier:2012ef, Blossier:2013ioa}:
\begin{align}
	\aT(k^2)&=\alpha(\zeta^2) k^2 \Delta(k^2;\zeta^2) F^2(k^2;\zeta^2),
	\label{aT}
\end{align}
where $\alpha(\zeta^2) = g^2(\zeta^2)/4\pi$; $F(k^2)$ is the ghost-propagator dressing function; and $L(k^2)$, which expresses additional aspects of ghost-gluon dynamics, satisfies a Dyson-Schwinger equation (DSE) [$\ell = k-q$]:
\begin{align}
L(k^2;\zeta^2) = g^2\!\! \int_{dq}^\Lambda  \left[4 \frac{(k \!\cdot\!q)^2}{k^2q^2}-1\right] B_1(q)\Delta(q^2;\zeta^2) \frac{F(\ell^2;\zeta^2)}{\ell^2}\,,
	\label{L-DSE}
\end{align}
with $B_1(q)$ being that single invariant in the ghost-gluon vertex which is nonzero in the limit of vanishing ghost momentum \cite{Aguilar:2013xqa, Dudal:2012zx}.\footnote{The expressions in Ref.\,\cite{Binosi:2014aea}, \emph{e.g}.\  Eq.\,(19), are recovered by using Eq.\,\eqref{aT} and recognising $F = 1/(1 + L + G)$ \cite{Grassi:2004yq, Aguilar:2009pp}.  Note, too, that herein we use $\Delta$ to express what is $\mathpzc{D}$ in Ref.\,\cite{Binosi:2014aea}.}

The gluon and ghost propagators are depicted in Fig.\,\ref{Delta-F}.  Their IR behaviour is controlled by the appearance of the gluon mass-scale, $m_g$, \emph{viz}.\ at O$(k^2)$ \cite{Boucaud:2010gr, RodriguezQuintero:2010wy, Aguilar:2013vaa, Athenodorou:2016oyh},
\begin{subequations}
\begin{align}
\label{gluonDelta}
\Delta^{-1}(k^2;\zeta^2) &\underset{k^2/\zeta^2\ll1}{\approx} k^2
\left(a_\Delta + \mathpzc{l}_g \ln{\frac{k^2+m_g^2}{\zeta^2}} + \mathpzc{l}_w \ln{\frac{k^2}{\zeta^2}} \right) + m_g^2, \\
F(k^2;\zeta^2) &\underset{k^2/\zeta^2\ll1}{\approx}  F(0;\zeta^2) \ \left( 1 + \frac{3}{16 \pi} \widehat{d}(0) \ k^2 \ln{\frac{k^2}{\zeta^2}}\right),
\label{DeltaF}
\end{align}
\end{subequations}
where $a_\Delta$, $\mathpzc{l}_g$, $\mathpzc{l}_w$ are simple constants.  Actually, inspection of Eq.\,\eqref{gluonDelta} reveals that $\mathpzc{l}_g$ and $\mathpzc{l}_w$, respectively, express the presence of massive-gluon and massless-ghost loops in the gluon vacuum polarisation.

The function $L$ is known to vanish at both IR and ultraviolet (UV) momenta \cite{Aguilar:2009nf}, and has the following soft-$k^2$ expansion:
\begin{align}
	L(k^2;\zeta^2) &\underset{k^2/\zeta^2\ll1}{\approx} - F^{-1}(0;\zeta^2)  \frac{\widehat{d}(0)}{4 \pi}  k^2\ \ln{\frac{k^2}{\zeta^2}}.
	\label{L}
\end{align}
It is implicit in Eq.\,\eqref{iKer} that the product $L F$ is RGI; and, indeed, using Eqs.\,\eqref{DeltaF}, \eqref{L}, one finds
\begin{subequations}
\label{LFINfIR}
\begin{align}
\label{LF}
	L(k^2) F(k^2) &\underset{k^2/\Lambda^2_\s{\mathrm{T}}\ll1}{\approx}  - \frac{\widehat{d}(0)}{4\pi} p^2 \ln{\frac{k^2}{\Lambda_\s{\mathrm{T}}^2}}, \\
	{\mathpzc I}(k^2)  &\underset{k^2/\Lambda^2_\s{\mathrm{T}}\ll1}{\approx}  k^2 \widehat{d}(0) \left[ 1 - \left( \frac{\widehat{d}(0)}{8\pi} + \frac{{\mathpzc l}_w}{m_g^2} \right) k^2\ln{\frac{k^2}{\Lambda_\s{\mathrm{T}}^2}} \right],
		\label{INfIR}
\end{align}	
\end{subequations}
where the renormalisation point $\zeta^2$ has been traded for $\Lambda^2_\s{\mathrm{T}}$, which is the textbook scale $\Lambda^2_\s{\mathrm{QCD}}$ evaluated within the Taylor scheme.   Eqs.\,\eqref{LFINfIR} emphasise the RGI character of $LF$ and $\mathpzc I$.

Eq.\,\eqref{INfIR} reveals a curious feature; namely, it directly connects the effect of massless-ghost loops in the gluon vacuum polarisation, typically identified solely with gluon-ghost dynamics, to the interaction strength which appears in the dressed-quark gap equation; and, hence, ultimately to quark confinement and DCSB.  Moreover, the expression of this connection is RGI because the ratio $\mathpzc{l}_w/m_g^2$ is independent of the renormalisation point.  Finally, a recent lQCD analysis of the three-gluon vertex indicates that $\mathpzc{l}_w>0$ \cite{Athenodorou:2016gsa}, which entails that massless-ghost loops enhance the IR strength of the gap equation's kernel. As will become apparent, this has important consequences.

Consider now the UV.  Owing to asymptotic freedom, UV dynamics is purely perturbative, and hence one can readily obtain
\begin{subequations}
\begin{align}
	k^2\Delta(k^2;\zeta^2) &\underset{k^2/\Lambda^2_\s{\mathrm{T}}\gg1}{\approx}  \left(\ln\frac{k^2}{\Lambda^2_\s{\mathrm{T}}} \Big/\ln\frac{\zeta^2}{\Lambda^2_\s{\mathrm{T}}}\right)^{-{\gamma_0}/{\beta_0}}, \\
	F(k^2;\zeta^2) &\underset{k^2/\Lambda^2_\s{\mathrm{T}}\gg1}{\approx}  \left(\ln\frac{k^2}{\Lambda^2_\s{\mathrm{T}}} \Big/\ln\frac{\zeta^2}{\Lambda^2_\s{\mathrm{T}}}\right)^{-{\widetilde{\gamma}_0/\beta_0}}, \\
	L(k^2;\zeta^2) &\underset{k^2/\Lambda^2_\s{\mathrm{T}}\gg1}{\approx}  \frac{3g^2(\zeta^2)}{32\pi^2}\left(\ln\frac{k^2}{\Lambda^2_\s{\mathrm{T}}} \Big/\ln\frac{\zeta^2}{\Lambda^2_\s{\mathrm{T}}}\right)^{-({\widetilde{\gamma}_0+\gamma_0)/\beta_0}},
\end{align}	
\end{subequations}
where $\gamma_0=13/2 - 2/3 n_f$, $\widetilde{\gamma}_0=9/4$, $\beta_0=11 -2/3 n_f$ are, respectively, the one-loop coefficients for the gluon and ghost propagator anomalous dimensions and the $\beta$-function.  Now, since $2\tilde{\gamma}_0+\gamma_0=\beta_0$, then:
\begin{subequations}
\begin{align}
	L(k^2) F(k^2) &\underset{q^2/\Lambda^2_\s{\mathrm{T}}\gg1}{\approx} \frac{3}{2\beta_0 \ln{(k^2/\Lambda_\s{\mathrm{T}}^2)}},\nonumber \\
	{\mathpzc I}(k^2)  &\underset{p^2/\Lambda_\s{\mathrm{T}}^2\gg1}{\approx} \alpha_\s{\mathrm{T}}(k^2) \underset{k^2/\Lambda_\s{\mathrm{T}}^2\gg1}{\approx} \frac{4 \pi}{\beta_0 \ln{({k^2}/{\Lambda^2_\s{\mathrm{T}}})}}.
	\label{INfUV}
\end{align}	
\end{subequations}

This analysis establishes that since the RGI product $LF$ vanishes both in the IR and the UV, then the gauge-sector interaction kernel, ${\mathpzc I}$, may only deviate from $\alpha_\s{\mathrm{T}}$ at intermediate momenta, in an amount controlled by the product $LF$ itself \cite{Aguilar:2009nf}.

\smallskip

\noindent\textbf{3.$\;$Scale setting}.  {\it Ab initio} evaluation of the interaction kernel in Eqs.\,\eqref{dkernel} requires knowledge of the gluon propagator $\Delta$, the ghost dressing function $F$ and the ghost-gluon form factor $B_1$, all renormalised at a certain scale $\zeta^2$ in the perturbative domain, \emph{i.e}.\ a scale at which the strong coupling can reliably be computed using perturbation theory.  Then, Eq.\,\eqref{aT} allows evaluation of $\aT$ and $L$ can be obtained by solving Eq.\,\eqref{L-DSE} using these inputs.

Employing this procedure, one could in principle use unquenched lQCD results for both two degenerate light flavours ($m_{u,d}\in[0.02,0.05]\,$GeV, quoted in the $\overline{\mathrm{MS}}$ scheme at $\zeta=2\,$GeV), and two degenerate light flavours plus two heavier quarks ($m_s = 0.095\,$GeV, $m_c=1.51\,$GeV) in order to estimate the response of the gauge-sector interaction kernel to the presence of $n_f$ dynamical quarks \cite{Ayala:2012pb}.  Care must be exercised, however, because if this path is followed in comparing quenched results \cite{Bogolubsky:2009dc} with $n_f=\Nft$, then one finds ${\mathpzc I}_{n_f=2}>{\mathpzc I}_{0}\sim{\mathpzc I}_{(2,1,1)}$; and hence, paradoxically, DCSB of greater strength in the presence of active, interaction-screening quarks than in their absence.

\begin{figure*}[!t]
	\begin{center}
	\hspace{-0.75cm}
	\includegraphics[width=0.475\linewidth]{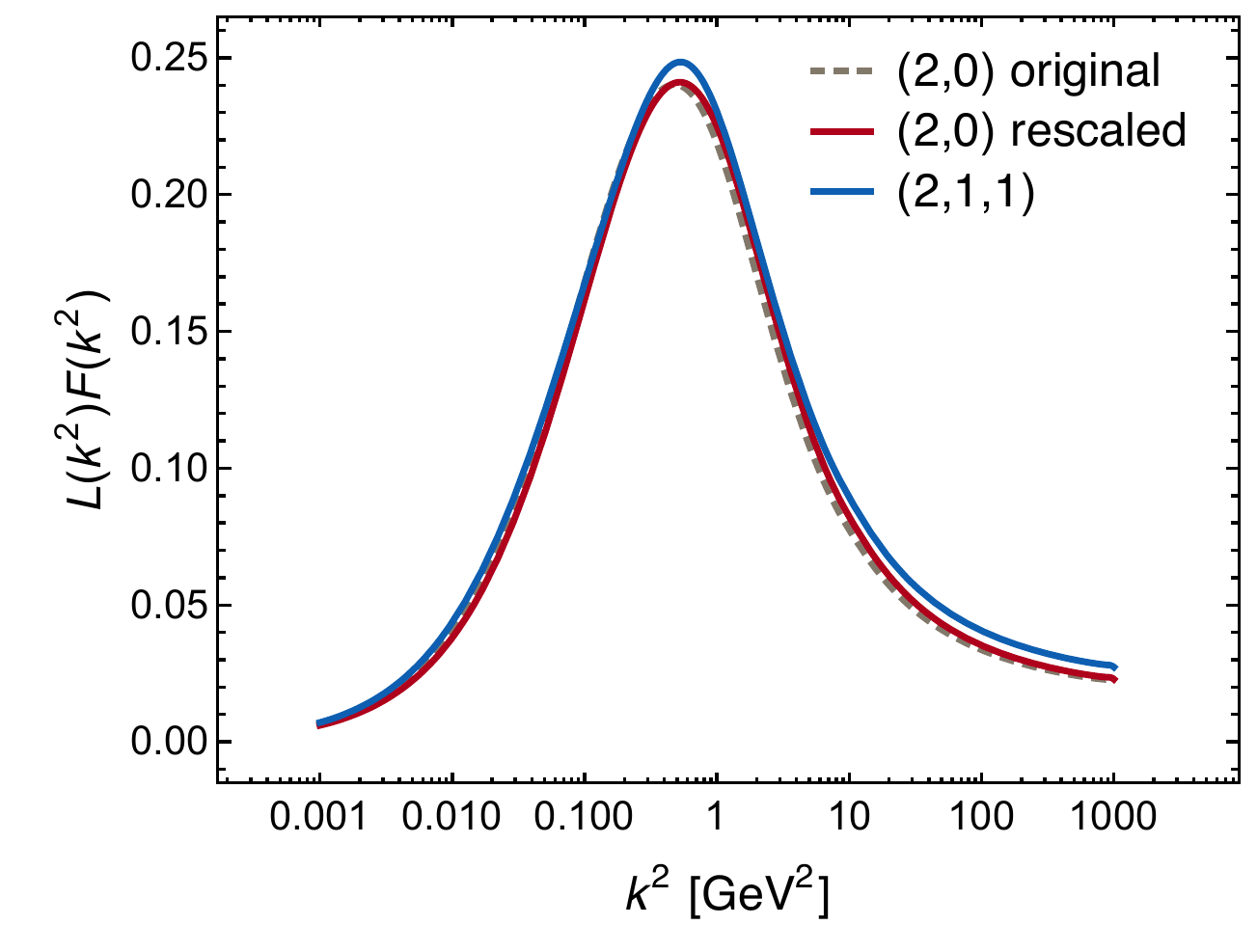}
	\hspace{0.5cm}
	\includegraphics[width=0.475\linewidth]{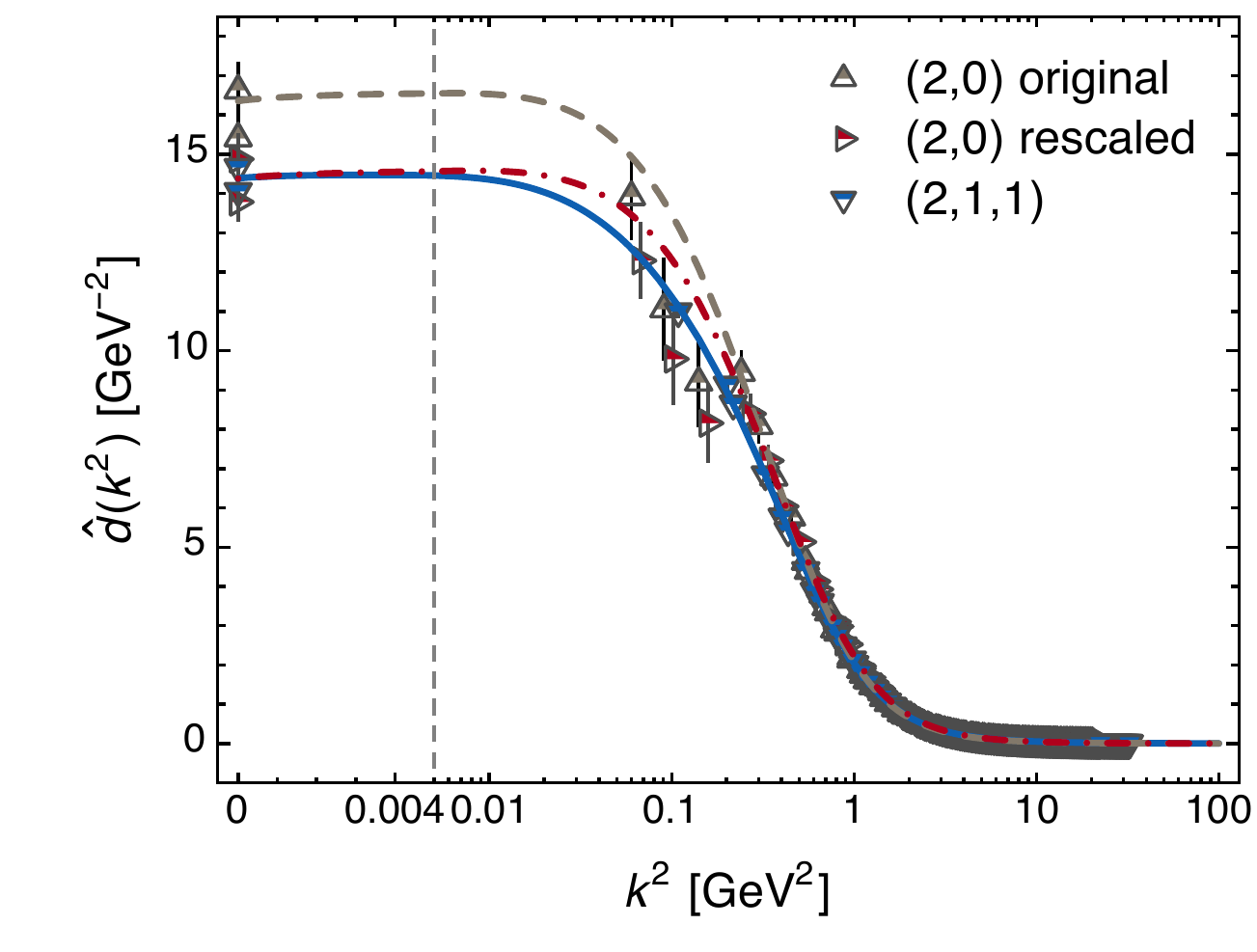}
	\end{center}
	\vspace{-.75cm}	
	\caption{\label{LF-dhat} RGI combinations entering the definition of the gauge-sector quark-gluon interaction kernel, Eqs.\,\eqref{dkernel}: $LF$ (left), and $\widehat{d}$ (right).  Plainly, using the original lQCD output: $\widehat{d}_{2+1+1}(0)\neq\widehat{d}_{2}(0)$; whereas the two curves almost overlap upon introduction of the rescaling factor in Eq.\,\eqref{rescfact}.
As in Fig.\,\ref{Delta-F}, the vertical dashed line in the right panel marks a change between linear and logarithmic scales for the $x$-axis.}
\end{figure*}

In order to understand this pathology, it is helpful to review the issue of scale setting.  In any lQCD simulation, the results are obtained in units specified by the lattice spacing, $a$, which is related to the lattice momentum via $q_\mu=(2\pi/a) l(\mu)/L(\mu)$, with $l(\mu)=1,\ldots,L(\mu)$ specifying the lattice site in the $\mu$ direction.  The physical magnitude of a given lattice momentum is therefore only determined once a relationship is drawn with some observable quantity.  This procedure is equivalent to fixing $\Lambda_\s{\mathrm{QCD}}$, the theory's fundamental RGI scale.  It is usually achieved by using lQCD to compute a specific reference quantity with the highest achievable precision and setting $a$ so that the computed result matches the empirical value.

It is immediately apparent, however, that such a procedure cannot be employed for quenched simulations: Nature offers no observable with which to compare; and, therefore, any choice is merely a theoretical convention.  Consequently, even supposing some array of quenched-lQCD Green functions match those of the corresponding Yang-Mills theory, and the running and effective couplings computed from three-point functions agree with perturbative calculations, no physical scale $\Lambda=1/a$ can meaningfully be inferred from these correspondences.

In unquenched simulations, on the other hand, a scale is typically chosen by fixing the pion's mass and leptonic decay constant.  This is valid for $n_f = \Nftpopo$.  However, systematic uncertainties, difficult to estimate, enter when the same is done for $n_f=\Nft$, since $s$-quarks do affect properties of light pseudoscalar mesons (\emph{e.g}.\ $\pi^0$-$\eta$ mixing \cite{Bhagwat:2007ha}), and only very careful and accurate accounting for such effects can enable a reliable determination of $a$ in this case.

\begin{table}[!t]
\begin{center}
\caption{\label{lambdaMSbar} Estimates of $\Lambda_{\overline{\s{\mathrm{MS}}}}$ extracted from the FLAG collaboration review \cite{Aoki:2016frl} and those inferred from experiments, corresponding to the PDG {\it world average} \cite{Agashe:2014kda}.  In the latter case, the two central values marked with an asterisk were obtained as explained in the text.}
\begin{tabular}{ccc}
\hline
{$n_f$}&\hspace{1cm} $\Lambda^\s{\mathrm{FLAG}}_{\overline{\s{\mathrm{MS}}}}$ [MeV]\hspace{1cm} & $\Lambda^\s{\mathrm{PDG}}_{\overline{\s{\mathrm{MS}}}}$ [MeV] \vspace{0.05cm}\\
\hline
0 & $260\ (7)$\;\; & $388^*\phantom{(10)}$  \\
$(2,0)$ & $330^{+21}_{-54}$\;\;\; & $364^*\phantom{(10)}$  \\
$(2,1,1)$ & $294\ (11)$ & $296\ (10)$ \\
\hline
\end{tabular}
\end{center}
\end{table}

The impact of $s$- and $c$-quarks on scale setting in lQCD is illustrated in Table~\ref{lambdaMSbar}, which reports estimates of $\Lambda_{\overline{\s{\mathrm{MS}}}}$ produced by the Flavor Averaging Lattice Group (FLAG) \cite{Aoki:2016frl} and the Particle Data Group (PDG) \cite{Agashe:2014kda}.  We will subsequently use $N_f$ to denote the number of light quarks, and $N_f^\prime$, $N_f^{\prime\prime}$, etc.\  to represent the number with given masses of increasing size.  With this notation, the PDG define the fundamental scale appropriate to $n_f=N_f+N_f^\prime+N_f^{\prime\prime} + \ldots$ with reference to $\Lambda_{\overline{\s{\mathrm{MS}}}}(n_f=5)$, \emph{viz}.\ $\Lambda_{\overline{\s{\mathrm{MS}}}}(n_f\neq 5)$ is the scale in an effective theory with $n_f\neq 5$ flavours which is tuned to describe observables at momenta that lie between the mass of the lighter $n_f-1$ quark and the heavier $n_f+1$ flavour.  Matching of the effective theories and determination of their respective scales is performed, implicitly, via the coupling itself, demanding
\begin{align}
	\alpha_{\overline{\s{\mathrm{MS}}}}^{n_f+1}(m_q) = \alpha_{\overline{\s{\mathrm{MS}}}}^{n_f}(m_q) \left\{ 1 + \sum_{i=1}^{i_m} c_{i0} \left[ \alpha_{\overline{\s{\mathrm{MS}}}}^{\nf}(m_q) \right]^i \right\},
\label{matching}
\end{align}
where $i_m=n-1$, when the running coupling is evaluated at $n$ loops, and $m_q$ defines the threshold of the $n_f+1$ quark flavor: $c_{10}=0$, $c_{20}=-11/[72\pi^2]$ ($\overline{\mathrm{MS}}$ scheme).  Additional details are presented elsewhere \cite{Agashe:2014kda}.

This prescription is consistent for all experiments at $\zeta \gtrsim m_c$, in which case $n_f=4=(2,1,1)$.  However, caution must be exercised when employing Eq.\,\eqref{matching} at the $s$-quark and lower thresholds, since they are located within the domain upon which non-perturbative effects influence the running of the coupling.  Acknowledging this difficulty, in Table~\ref{lambdaMSbar} we define ``PDG-like'' values of $\Lambda_{\overline{\s{\mathrm{MS}}}}$ for the $n_f=\Nft$ and quenched cases by using Eq.\,\eqref{matching} with $u/d$- and $s$-quark thresholds located at $1\,$GeV, \emph{i.e}.\ approximately the proton mass, a natural scale for light-quark physics.

This discussion highlights that values of the lattice spacing, $a$, for quenched and $n_f=\Nft$ simulations are typically not set realistically in order to account for the decoupling of $u/d$- and $s$-quarks from, \emph{e.g}.\ the $n_f=\Nftpopo$ theory.  We cannot fix the quenched case; but, as explained below, a procedure does exist which can be used to produce a valid value of $a$ for $n_f=\Nft$.

The natural requirement that heavy species ($N_f^\prime$, $N_f^{\prime\prime}$,...) decouple from light ones ($N_f$) after crossing the corresponding thresholds, implies the interaction kernel should be such that
\begin{align}
	&\lim_{k^2\to0}\frac{{\mathpzc I}_{n_f}(k^2)}{k^2}=\lim_{k^2\to0}\frac{{\mathpzc I}_{N_f}(k^2)}{k^2}& &\Leftrightarrow & \widehat{d}_{n_f}(0)=\widehat{d}_{N_f}(0).&
	\label{res-cond}
\end{align}
This condition can in turn be used to fix the fundamental scale of the $N_f$ theory in terms of the $n_f$ theory, which, containing heavier quarks, is implicitly assumed to more accurately capture QCD's dynamics.  Eq.\,\eqref{res-cond} cannot, however, be used for setting the quenched scale from that appropriate to $n_f=\Nft$ because in the latter case the chiral limit is usually used for the scale setting \cite{Blossier:2010ky}, whereas the quenched case corresponds to static (infinitely massive) quarks.

\begin{figure*}[!t]
	\begin{center}
	\hspace{-0.75cm}
	\includegraphics[width=0.475\linewidth]{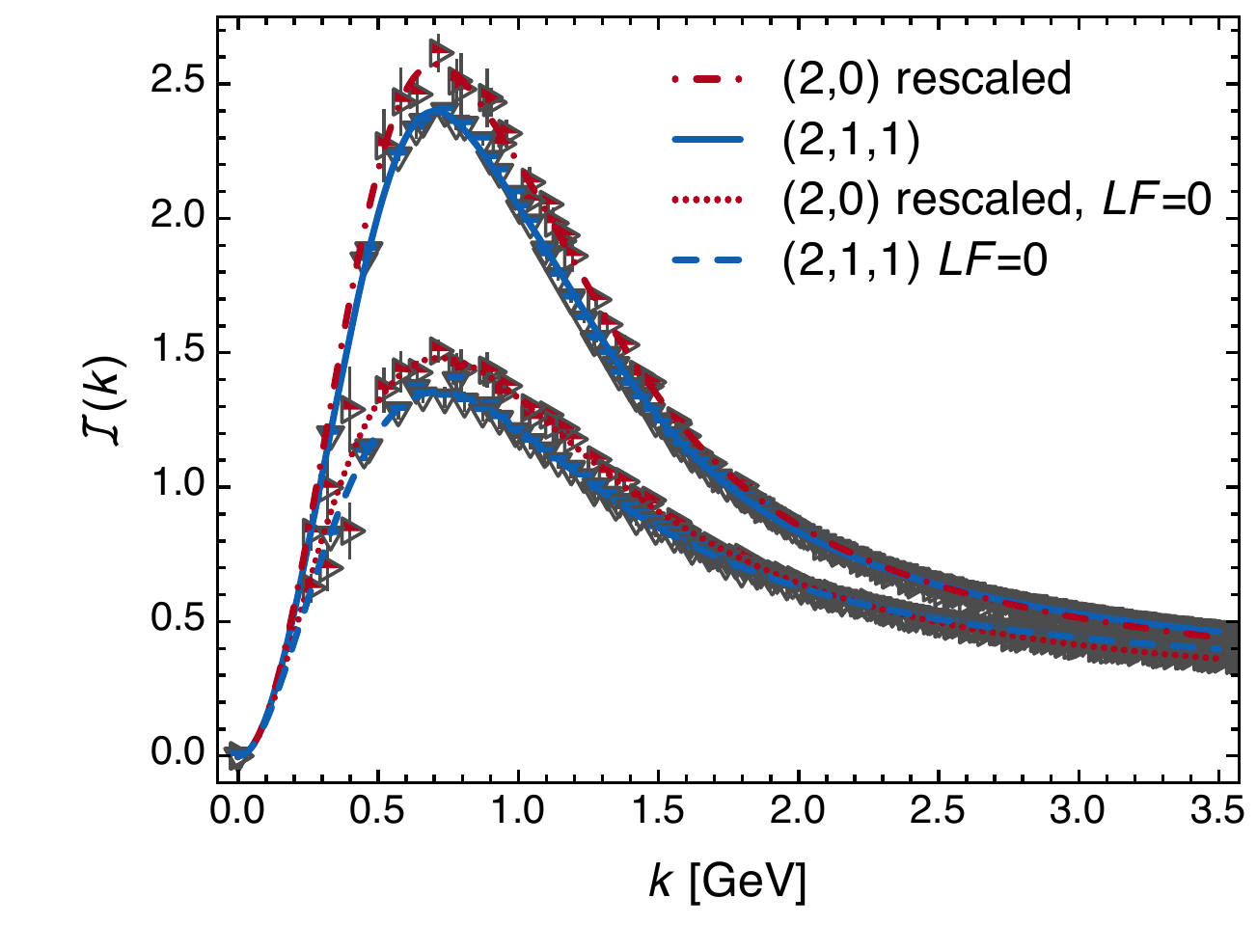}   
	\hspace{0.5cm}
	\includegraphics[width=0.475\linewidth]{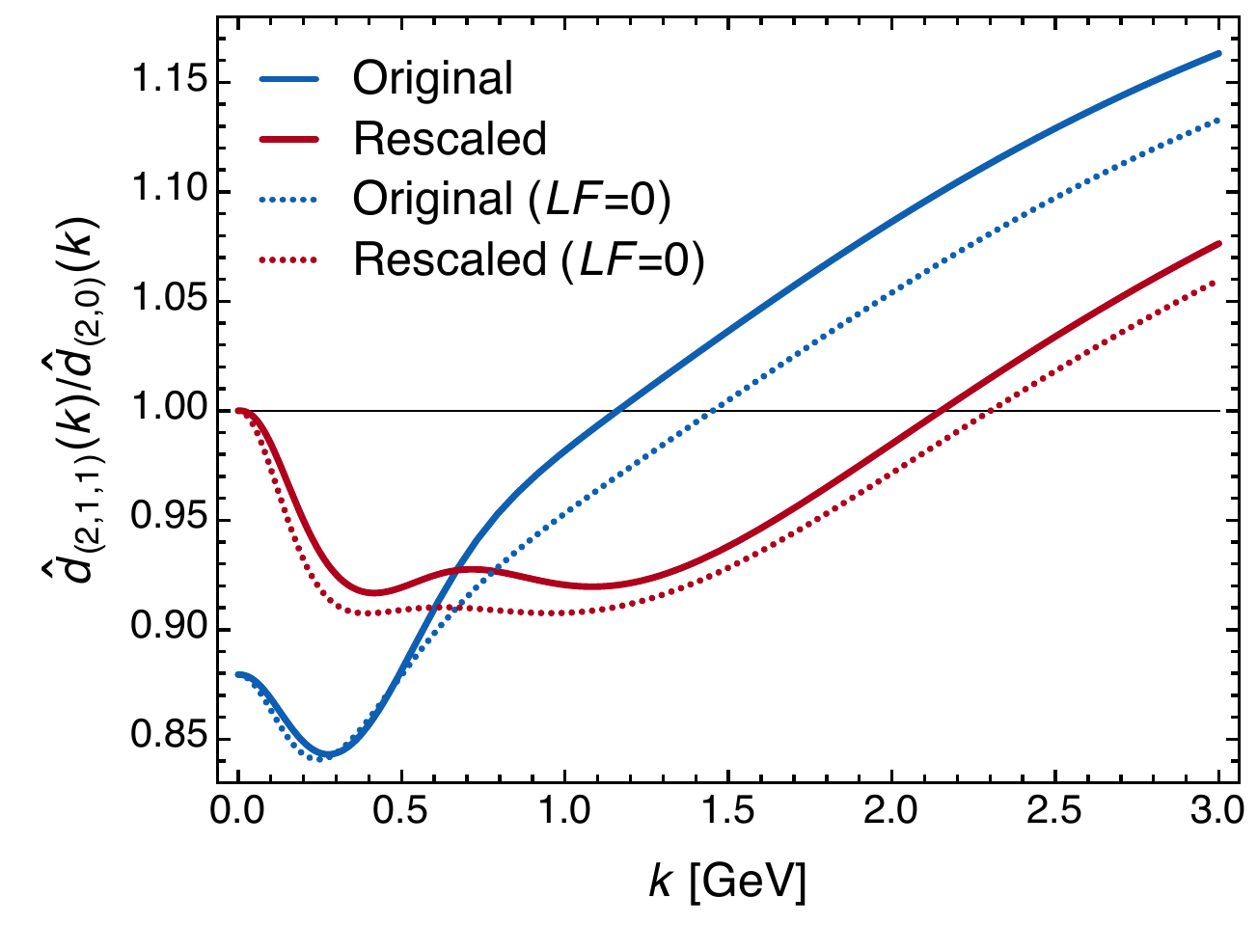}  
	\end{center}
	\vspace{-0.75cm}
	\caption{\label{INf} (Left panel) The interaction strength for $\Nft$ and $\Nftpopo$ (continuous lines) and the corresponding $\alpha_\s{\mathrm{T}}$ (dashed lines).  (Right panel) The ratios ${\widehat d}_{2+1+1}/{\widehat d}_2$ (continuous lines) and the corresponding ratio of the Taylor coupling (dashed lines) using the original and rescaled $\Nft$ data.}
\end{figure*}

Using lQCD results \cite{Ayala:2012pb} for the gluon propagator and ghost dressing function with $n_f=\Nft$, $\Nftpopo$, as in Fig.\,\ref{Delta-F}, one can construct the RGI combination $\widehat d$ in both cases.  (\emph{N.B}.\ $\alpha=0.33$ and 0.37 at $\zeta_4$ for $\Nft$ and $\Nftpopo$, respectively \cite{Blossier:2010ky, Blossier:2012ef}.) As evident in the right panels of Figs.\,\ref{LF-dhat} and Fig.\,\ref{INf}, however, with the value of $a$ determined in Ref.\,\cite{Ayala:2012pb}, Eq.\,\eqref{res-cond} is violated.
Demanding, on the other hand, that Eq.\,\eqref{res-cond} is fulfilled, one is led to introduce a rescaling factor:
\begin{align}
	{\mathpzc s}_a=\sqrt{\frac{\widehat{d}_{(2,0)}(0)}{\widehat{d}_{(2,1,1)}(0)}}
	\label{rescfact}
\end{align}
and a new, correlated lattice spacing $a^\prime = a/{\mathpzc s}_a$, in terms of which all quantities associated with the $n_f=\Nft$ configurations should be recomputed.  Specifically, the corrected value of a given quantity at momentum $q$ is equal to the original value determined at $q/{\mathpzc s}_a$:
\begin{equation}
P_{\rm corrected}(q) = P_{\rm original}(q/{\mathpzc s}_a) \,.
\label{rscale}
\end{equation}
Applying Eq.\,\eqref{rescfact} to the results in Ref.\,\cite{Ayala:2012pb}, one obtains ${\mathpzc s}_a=1.06$, leading to the rescaled gauge-sector functions depicted in Fig.\,\ref{Delta-F} and a new $n_f=\Nft$ value of $\alpha=0.35$ at $\zeta_4$.

The rescaled running interaction is depicted in the left panel of Fig.\,\ref{INf}.  In accordance with physics-based expectations, at momenta far below the $s$- and $c$-quark thresholds, ${\mathpzc I}_{\Nft} = {\mathpzc I}_{\Nftpopo}$; at larger momenta, still below roughly 2\,GeV, ${\mathpzc I}_{\Nftpopo} < {\mathpzc I}_{\Nft}$; and, finally, on the remaining spacelike domain, the hierarchy is inverted, with ${\mathpzc I}_{\Nftpopo} > {\mathpzc I}_{\Nft}$ simply because the Taylor coupling's perturbative $\beta$-function decreases as the number of active quark flavour increases.
These features are also evident in the ratios drawn in the right panel of Fig.\,\ref{INf}, which highlight the suppression of ${\widehat d}_{\Nftpopo}$ with respect to ${\widehat d}_{\Nft}$ on that domain of momenta which contains the heavier-quark thresholds.
Curiously, when using our rescaling factor, the FLAG $n_f=\Nft$ estimate in Table~\ref{lambdaMSbar} changes to $\Lambda_\s{\overline{\mathrm{MS}}}=0.350^{+22}_{-57}$ MeV, thereby becoming compatible with our estimate for the PDG value, obtained using Eq.\,\eqref{matching} with a $n_f=3$ threshold located at 1\,GeV.

\smallskip

\noindent\textbf{4.$\;$Flavor dependence of the interaction}.
We have developed an interpolation that describes the curves in Fig.\,\ref{INf}, preserving the IR and UV behaviour presented in Eqs.\,\eqref{INfIR}, \eqref{INfUV}, \emph{viz}.\
\begin{align}
	\widehat{d}(k^2)
= \widehat{d}(0) & \ \frac{\displaystyle 1 - d_1 k^2\ln[1+\Lambda_0^2/k^2] + a_1 k^2}{1+b_1 k^2 + b_2 k^4 + b_3 k^6} \nonumber \\
	&+ \frac{4\pi k^4}{\beta_0 \left( \Lambda_0^6+k^6 \ln{\frac{k^2}{\Lambda_T^2}}\right)},
\label{ansatz}
\end{align}
with the coefficients listed in Table~\ref{params}.
Expanding the interpolation to ${\mathpzc O}(k^2)$ and comparing the result with Eq.\,\eqref{INfIR}, one finds
\mbox{$-d_1=\widehat{d}(0)/(8\pi)+{\mathpzc l}_w/m_g^2$}; and substituting the values of $d_1$ in Table~\ref{params}, one obtains ${\mathpzc l}_w/m_g^2 = 1.71\,$GeV$^{-2}$ for both $\Nft$ and $\Nftpopo$.
This may be understood by recalling that ${\mathpzc l}_w$ is generated by massless-ghost loops and should therefore be rather insensitive to the number of quarks; and $m_g^2$ defines the $k^2=0$ value of the gluon propagator, which cannot sensibly depend on the number of heavy (inactive) quarks.

\begin{table}[!b]
\begin{center}
\caption{\label{params} Interpolation coefficients in Eq.\,\eqref{ansatz}, relating to the interaction kernels obtained with $n_f=\Nft$ (rescaled) and $n_f=\Nftpopo$.  They carry mass-dimension: GeV$^{-2}$ for $\widehat d(0)$; and GeV$^{-2i}$ for those quantities with subscript $i\in \mathbb{N}$.   We have fixed $\Lambda_T=0.5\,$GeV and $\Lambda_0=1\,$GeV.
}
\begin{tabular}{ccccccc}
\hline
\hline
$n_f$  & $d_1$ & $a_1$ & $b_1$ & $b_2$ & $b_3$ & $\widehat{d}(0)$ \\
\hline
(2,0) & -2.276 & 1.809 & \,\;9.93 & 1.100 &  22.41 & 14.38\\
(2,1,1) & -2.289 & 1.518 & 11.72 & -3.864 & 28.02 & 14.38 \\
\hline
\end{tabular}
\end{center}
\end{table}

\begin{figure*}[!t]
\hspace*{\fill}%
\begin{minipage}[t]{0.5\textwidth}
\centering
\vspace{0pt}
\includegraphics[width=0.95\textwidth]{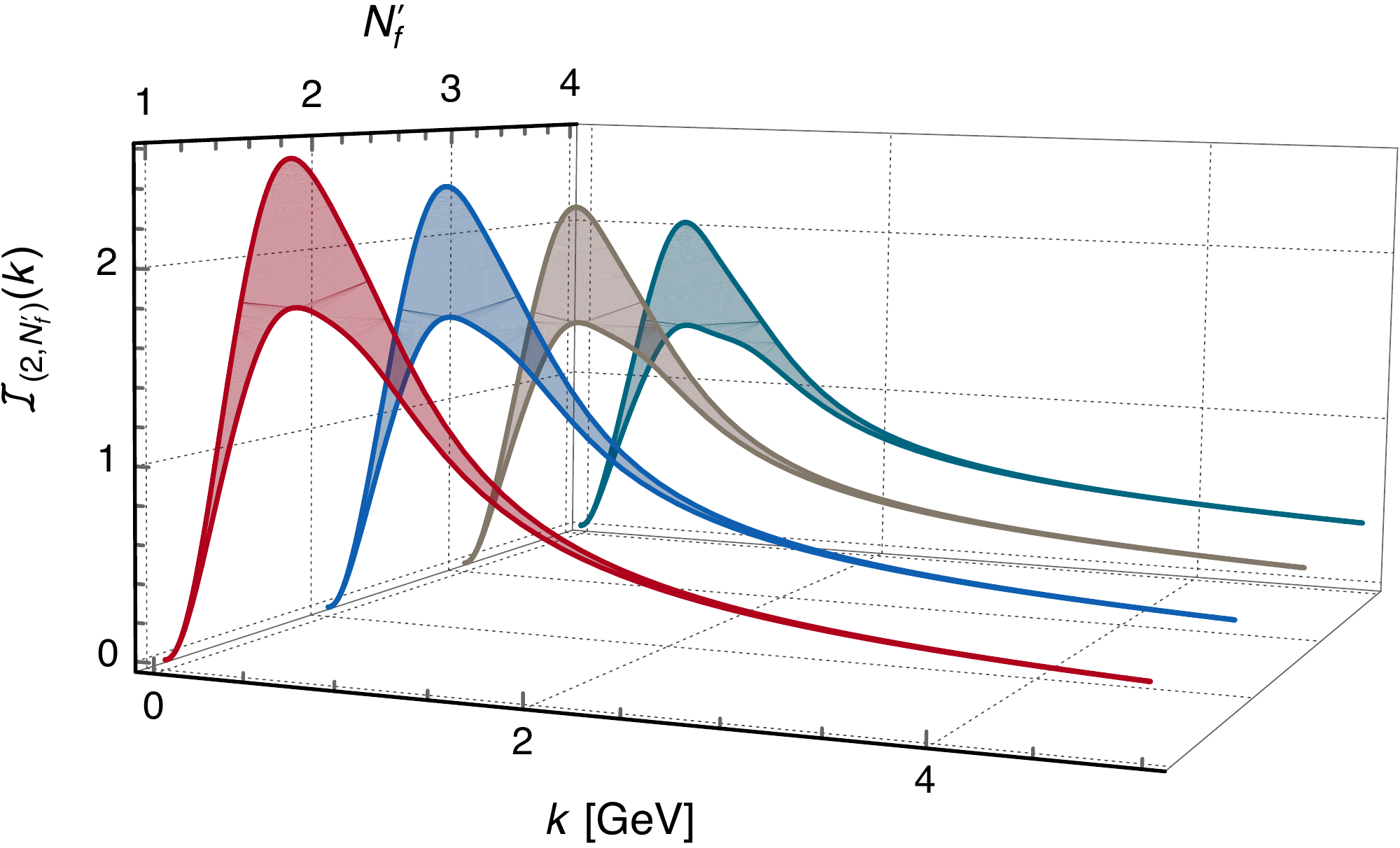}    
\end{minipage}%
\hfill
\begin{minipage}[t]{0.5\textwidth}
\centering
\vspace{0pt}
\includegraphics[width=0.95\textwidth]{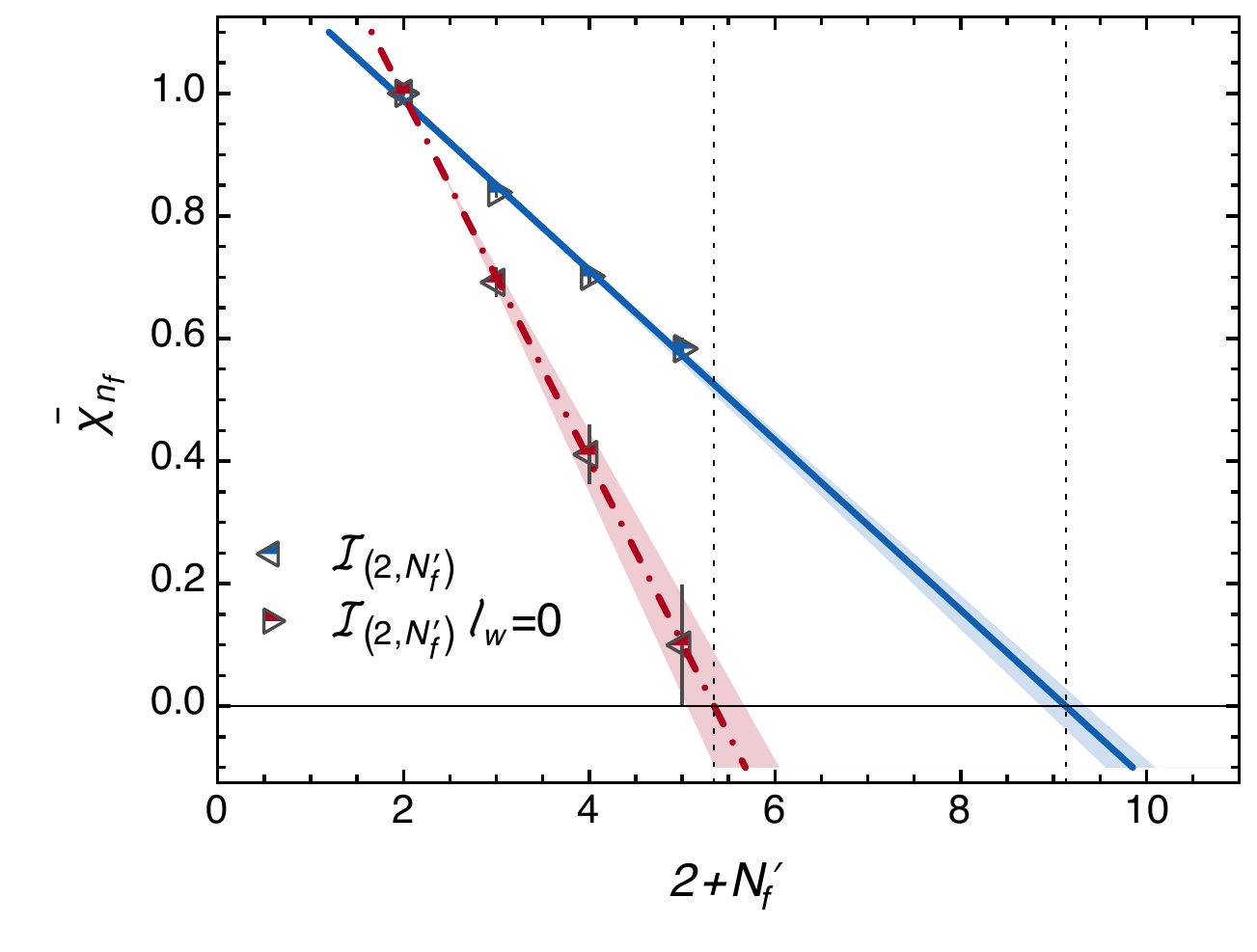}   
\end{minipage}%
\hspace*{\fill}
	\caption{\label{deltaNf}
(Left panel) Gap equation RGI interaction kernel (upper curves) for a $(2,N_f^\prime)$ theory with $N_f^\prime =0,\dots,3$.  The coefficients in Eq.\,\eqref{ansatz} are: $a_1=1.81- 0.292\,N_f^\prime$, $b_1=9.93+1.79\,N_f^\prime$, $b_2=1.1-4.96\,N_f^\prime$, $b_3=22.41+5.61\,N_f^\prime$.  The corresponding kernels obtained with ${\mathpzc l}_f=0$ are also depicted (lower curves).
(Right panel) Chiral order parameter in Eq.\,\eqref{chimean}, which exposes the impact on DCSB of adding additional $s$-quark-like active quarks to the theory.  Extrapolating linearly, DCSB is absent in this class of theories for $n_f =2+N_f^\prime \gtrsim 9$ ($n_f \gtrsim 5$ in the absence of massless-ghost loops).}
\end{figure*}

The parametrisation in Eq.\,\eqref{ansatz} enables us to sketch the dependence of $\widehat{d}(k^2)$ on the number of active quarks.  To proceed, we note that the largest part of the non-perturbative difference between ${\mathpzc I}_{\Nftpopo}$ and ${\mathpzc I}_{\Nft}$ is located below the $c$-quark threshold (see Fig.\,\ref{INf}) and therefore assume that it can largely be attributed to the $s$-quark, $m_s=95\,$MeV, \emph{i.e}.\ $\Nftpopo \approx (2,1)$.  This deduced, then the coefficients in Eq.\,\eqref{ansatz} can be related thus:
\begin{equation}
d_1^{\Nftpopo} = d_1^{\Nft} + \delta_{N_f^\prime}\, \delta d_1\,,
\label{deltad}
\end{equation}
etc., where $\delta d_1 = d_1^{\Nftpopo}-d_1^{\Nft}$ and $\delta_{N_f^\prime} =1$ because a single active $s$-quark-like flavour has been added.
We next assume, too, that Eq.\,\eqref{deltad}, and its partners for the other coefficients, can serve unchanged on $\delta_{N_f^\prime} \geq 2$.
These two assumptions yield the running interactions for theories with $n_f = (2,N_f^\prime)$, $N_f^\prime=1,2,3$, depicted in the left panel of Fig.\,\ref{deltaNf}.  Results obtained in the absence of ghost-loop enhancement (${\mathpzc l}_w=0$) are also drawn.  Plainly, massless-ghost loops significantly enhance the interaction strength at IR momenta, a result telegraphed by the fact that ${\mathpzc l}_w/m_g^2 \approx  3\times \widehat d(0)/(8\pi)$, as pointed out following Eq.\,\eqref{ansatz}.

\smallskip

\noindent\textbf{5.$\;$Chiral symmetry restoration}.
%
%
We are now in a position to combine all features of the preceding discussion and explore the impact on DCSB of adding active quark flavours to the theory: $n_f=(2,0) \to (2,N_f^\prime)$, addressing the question of whether there is a critical number, $n_f^{\rm cr} = 2+N_f^{\rm cr}$, above which the interaction cannot support DCSB.
For the answer to be reliable, however, a realistic dressed--gluon-quark vertex, $\Gamma_\nu$, must be employed in the gap equation, Eq.\,\eqref{gendseN}, because positive feedback introduced by that vertex is known to enhance DCSB and, indeed, without it, reconciliation of the top-down and bottom-up approaches to determining QCD's RGI running interaction is impossible \cite{Binosi:2014aea}.

In principle, the strong-interaction sector of the Standard Model is characterised by a unique form of $\Gamma_\nu$.  That form is not yet known, but recent work \cite{Binosi:2016wcx} has severely limited the class of realistic \emph{Ans\"atze} by using just three physical constraints.  This class may be expressed as follows $(t=q+p)$:
\begin{subequations}
\begin{align}
\Gamma_\nu(q,p) & = \Gamma_\nu^{\rm BC} (q,p) + \Gamma_\nu^{\rm T}(q,p),\\
\label{BCvertex}
i \Gamma_\nu^{\rm BC} (q,p) & = i \gamma_\nu \Sigma_A^{qp}
+ t_\nu [ i \tfrac{1}{2} \gamma\spr t\, \Delta_A^{qp} + \Delta_B^{qp} ], \\
\label{Tvertex}
\Gamma_\nu^{\rm T}(q,p) & =  \frac12 t_\nu^\s{\mathrm{T}} \sigma_{\alpha\beta} q_\alpha p_\beta\, \tau_4^{qp}+\sigma_{\nu\rho}\, (q-p)_\rho\tau_5^{qk}\nonumber \\
	&+(q_\nu \gamma\spr p - p_\nu \gamma\spr q + i \gamma_\nu \,\sigma_{\alpha\beta} \,q_\alpha p_\beta)\tau_8^{qk},
\end{align}
\end{subequations}
where $\lambda_1^{qp} = \Sigma_A^{qp} = [A(q^2)+A(p^2)]/2$, $\lambda_2^{qp}= \Delta_A^{qp}$,  $\lambda_3^{qp}= \Delta_B^{qp}$,  $\Delta_\phi^{qp}=[\phi(q^2)-\phi(p^2)]/[q^2-p^2]$, $\phi=A,B$, $t_\nu ^\s{\mathrm{T}}=T_{\nu\rho}t_\rho$; with
\begin{align}
   \tau_4^{qk} & = a_4 \frac{4 \Delta_B^{qk}}{t^\s{\mathrm{T}}\spr t^\s{\mathrm{T}}};&
	\tau_5^{qk}  &= a_5 \Delta_B^{qk};&
 	\tau_8^{qk}  &= a_8 \Delta_A^{qk},
\end{align}
where $a_{4,5,8}$ are dimensionless constants modulating the strength of the associated vertex term. Simple algebra shows that the gap equation's kernel does not depend separately on $a_{4}$, $a_{5}$, but, instead, only on the combination $a_{\widehat{45}} = a_4-3a_5$; and the class of realistic \emph{Ans\"atze} is then specified by the domain\footnote{%
A larger class of \emph{Ans\"atze} was identified in Ref.\,\cite{Binosi:2016wcx}, involving two additional Dirac-matrix structures and hence two more coefficients.  However, $a_{\widehat{45}}$, $a_8$ and the associated tensors are by far the most important in connection with DCSB, and that is why we simplify the form.  Consequently, $\mathbb{V}_2\subset \mathbb{G}_4$, where $\mathbb{G}_4$ is the extremely small subdomain of $\mathbb{R}^4$ that contains all acceptable \emph{Ans\"atze}.}
\begin{align}
	\mathbb{V}_2=\{(a_{\widehat{45}},a_8)\,|\,	a_{\widehat{45}}\in[-0.95,-0.7],a_8\in[-1.3,-0.73]\}\,.
\end{align}
The class of \emph{Ans\"atze} thus defined involves only those functions that appear in the quark propagator, and hence its $n_f$-dependence is completely specified by the analysis in Sec.\,4.

At this point, consider a vertex
$^\mathfrak{q}\Gamma_\nu$, where \mbox{$\mathfrak{q} = (a_{\widehat{45}},a_8)$} is a vector in $\mathbb{V}_2$.  For a given value of $n_f$, we solve the chiral-limit gap equation for every such vertex $^\mathfrak{q}\Gamma_\nu$ identified in Ref.\,\cite{Binosi:2016wcx} using the RGI running interaction, ${\mathpzc I}_{(2,N_f^\prime)}$, described in Sec.\,4.  From the associated solutions, we construct the RGI ratio
$^\mathfrak{q}M_{(2,N_f^\prime)}(p^2)= \mbox{}^\mathfrak{q}B_{(2,N_f^\prime)}(p^2)/ \mbox{}^\mathfrak{q}A_{(2,n_f^\prime)}(p^2)$ for $N_f^\prime =0,\dots,3$,
and subsequently
$\mbox{}^\mathfrak{q}M_{(2,N_f^\prime)}(0)/\mbox{}^\mathfrak{q}M_{(2,0)}(0)$,
which measures the impact of an increasing number of active $s$-quark-like flavours on the existence and strength of DCSB.
Finally, we average the results over $\mathfrak{q}\in \mathbb{V}_2$ to obtain
\begin{equation}
\label{chimean}
\bar\chi_{n_f} := {\rm Mean}_{\mathfrak{q}\in \mathbb{V}_2}[M_{(2,N_f^\prime)}(0)/M_{(2,0)}(0)]\,,
\end{equation}
identifying the standard-deviation as the statistical error.

The outcome of this procedure is depicted in the right panel of Fig.\,\ref{deltaNf}.  The triangles indicate results from our direct calculations, whereas the lines are linear interpolations.  The evident accuracy of those interpolations encourages us to infer the existence and location of a critical number of flavours by extrapolation; and we thereby find that in a theory with $n_f=2+N_f^\prime$, \emph{i.e}. 2 light quarks and $N_f^\prime$ active $s$-like quarks, DCSB is impossible for $n_f > n_f^{\rm cr}$, where
\begin{align}
\label{nfcr}
n_f^{\rm cr} & = 2 + N_f^{\prime\,{\rm cr}} = 9.1 \pm 0.3\,.  
%
\end{align}
On the other hand, if one omits the enhancement generated by massless-ghost loops, setting ${\mathpzc l}_w = 0$, then $n_f^{\rm cr}= 5.4 \pm 0.3$.

In order to provide a context for the critical value in Eq.\,\eqref{nfcr}, we note that numerous lQCD analyses have attempted to address the same problem, finding a value of $n_f^{\rm cr}$ that lies somewhere between $n_f=8$ and $n_f=10$ \cite{Appelquist:2009ka, Hayakawa:2010yn, Cheng:2013eu, Aoki:2013xza}.   The evident agreement is meaningful because the approaches are so completely different.
We analyse the RGI gauge-sector running interaction, advocate a physical scale-setting procedure, subsequently infer the interaction's evolution with increasing numbers of active $s$-like quarks, and finally solve the gap equation in the chiral limit using the flavour-dependent interaction and a class of modern \emph{Ans\"atze} for the dressed--gluon-quark vertex.
On the other hand, one way or another, lQCD simulations explicitly break chiral symmetry; and, consequently, all chiral symmetry order parameters are necessarily nonzero.  A given order parameter (or collection thereof) is nevertheless computed on such configurations, and its dependence on the number of dynamical quarks is measured for some small number of current-quark masses.  Finally, a chiral extrapolation is performed in order to obtain and estimate for $n_f^{\rm cr}$ in the chiral limit.  The issue of scale setting (associating a physically meaningful value to the lattice spacing, $a$) is also important here because Nature does not provide empirically accessible examples with zero, or six, eight, ten light quarks.  This attaches additional uncertainty to the chiral extrapolation because one cannot be certain that all or even some of the input masses of the dynamical quarks actually lie within a domain that allows a reliable extrapolation.
Notwithstanding the vast differences in method, our result and those from lQCD agree within 10\%, an outcome which boosts confidence in the possibility that QCD with $n_f\lesssim n_f^{\rm cr}$ is a conformal theory.

%
It is incumbent upon us here to remark upon the semi-quantitative agreement between the result in Ref.\,\cite{Bashir:2013zha}, $n_f^{\rm cr} \sim 8 \pm 1$, and ours, Eq.\,\eqref{nfcr}.  
Ref.\,\cite{Bashir:2013zha} did not incorporate the necessary rescaling of the interaction discussed herein and employed a tree-level gluon-quark vertex, investing all the strength needed for DCSB at the empirical value $n_f=(2,1)$ in an over-amplification of the ``effective interaction'' at IR momenta.  It is therefore largely lacking in the connections with QCD that our analysis maintains.  
On the other hand, the foundation for Ref.\,\cite{Bashir:2013zha} is a model interaction \cite{Maris:1997tm, Maris:1999nt, Qin:2011dd} tuned to achieve a good description of in-vacuum light-quark observables when used with the leading-order (rainbow-ladder) truncation \cite{Binosi:2016rxz} of the strong-interaction's matter-sector DSEs; and the study explored the effect of two vastly different assumptions about the flavour-dependence of that interaction, thus determining $7\lesssim n_f^{\rm cr} \lesssim 9$.  In following this path, Ref.\,\cite{Bashir:2013zha} provided a well-motivated projection and a sensible error estimate.
It is worth noting that if we were to employ the deconfinement criterion exploited in Ref.\,\cite{Bashir:2013zha}, then we would find that quark confinement is also lost when $n_f$ exceeds $n_f^{\rm cr}$ in Eq.\,\eqref{nfcr}.

\smallskip

\noindent\textbf{6.$\;$Conclusion}.
%
We extended the renormalisation-group-invariant (RGI) running interaction computed in an \emph{ab initio} analysis of quenched gluon-ghost dynamics, incorporating effects generated by a number of light- and heavy-quark flavours [Fig.\,\ref{deltaNf}].  The sole inputs were results from unquenched lattice-QCD (lQCD) studies of the theory's gauge-sector two-point functions.
Our analysis revealed a systematic error in the procedure used to set the lattice scale in simulations of Yang-Mills theories whose flavour content is not precisely that of QCD, and we proposed a way to eliminate it [Eqs.\,\eqref{res-cond}\,--\,\eqref{rscale}].
These advances enabled us to introduce a parametrisation of the running interaction [Eq.\,\eqref{ansatz}, Fig.\,\ref{deltaNf}], which respects its model-independent infrared and ultraviolet behaviours, and simultaneously expresses its dependence on the number $n_f=2+N_f^\prime$, $N_f^\prime=1,\ldots,3$, of active quarks: $u$, $d$, and $N_f$ $s$-like quarks.

Using this RGI running interaction in concert with a well constrained class of dressed--gluon-quark vertices, we estimated the critical number of active lighter-quarks above which DCSB becomes impossible: $n_f^{\rm cr}=2+N_f^{\prime{\rm cr}}\approx 9$.
A particular qualitative feature of our analysis is the manner by which it draws a direct connection between the action of massless-ghost loops in QCD's gauge sector and measurable hadron properties, \emph{e.g}.\ such loops are responsible for an enhancement of the running interaction at intermediate momenta, critical to DCSB, and they are also the origin of a zero and subsequently a logarithmic divergence in some of the coefficient functions that characterise the dressed--three-gluon vertex.  In the absence of such loops, $n_f^{\rm cr}=2+N_f^{\prime{\rm cr}}\approx 5$, which is physically untenable.

\smallskip

\noindent\textbf{Acknowledgments}.
We are grateful for useful input from S.-X.~Qin.
Completion of this research was facilitated by ``NpQCD16'', the $3^{rd}$ Workshop on Non-perturbative QCD, University of Seville, Seville, Spain, 17-21 October 2016.
The results described herein were obtained using the KORE HPC of the Fondazione Bruno Kessler.
This research was supported by:
U.S.\ Department of Energy, Office of Science, Office of Nuclear Physics, contract no.~DE-AC02-06CH11357;
and Spanish MEYC under grant FPA2014-53631-C-2-P.



\end{document}